\documentclass[article]{elsarticle}
\usepackage{lineno,float,verbatim,color}
\usepackage[margin=3.8cm]{geometry}
\newcommand\T{\rule{0pt}{2.6ex}}
\newcommand\B{\rule[-1.2ex]{0pt}{0pt}}
\usepackage[colorlinks]{hyperref}
\usepackage{amsmath,amsfonts,amssymb}

\journal{Nuclear Physics B}
\begin{document}
\begin{frontmatter}

\title{Limits on new coloured fermions using precision \\ jet data from the Large Hadron Collider}

\author[1]{Javier Llorente}
\ead{javier.llorente.merino@cern.ch}
\address[1]{Institute of High Energy Physics (Chinese Academy of Sciences)}

\author[2]{Benjamin P. Nachman}
\ead{bnachman@cern.ch}
\address[2]{Lawrence Berkeley National Laboratory}

\begin{abstract}
This work presents an interpretation of high precision jet data from the ATLAS experiment in terms of exclusion limits for new coloured matter. To this end, the effect of a new coloured fermion with a mass $m_X$ on the solution of the renormalization group equation QCD is studied. Theoretical predictions for the transverse energy-energy correlation function and its asymmetry are obtained with such a modified solution and, from the comparison to data, 95\% CL exclusion limits are set on such models.
\end{abstract}

\end{frontmatter}


\section{Introduction}
The discovery of the Higgs boson by the ATLAS~\cite{Aad:2012tfa} and CMS experiments~\cite{Chatrchyan:2012xdj} in 2012 marked the completion of the Standard Model (SM) of particle physics.  However, even with direct measurements of all SM parameters, it is not possible to predict all known phenomena as the SM does not describe dark matter nor is it a quantum theory of gravity.  These challenges, in addition to other technical and aesthetic ones, have motivated an impressive program to search directly for new particles and forces at the Large Hadron Collider (LHC)~\cite{atlassusytwiki,atlasexoticstwiki,cmsexoticstwiki,cmssusytwiki,cmsb2gtwiki}.  Despite the lack of significant evidence for physics beyond the SM, direct searches will continue to play a central role in the LHC program.  

An important complement of direct searches is indirect searches by testing the consistency of the SM.  Recent global fits to electroweak observables are able to place important constraints on new phenomena, in particular related to Higgs bosons~\cite{Ellis:2018gqa,Haller:2018nnx}.  The strong force sector of the SM can also be used to set (nearly) model-independent constraints on new colored particles.  In particular, measurements sensitive to hard quantum chromodynamical (QCD) radiation at multiple scales can probe physics beyond the SM (BSM) through the running of the strong coupling constant, $\alpha_s$.   For example, an interpretation of $e^+e^-$ measurements in Ref.~\cite{kaplan} sets robust limits on gluinos up to about 50 GeV.  A similar strategy with $pp$ data in Ref.~\cite{spannowsky} extends these limits to about 300 GeV.  Direct searches for gluinos are currently able to probe the multi-TeV regime~\cite{atlassusytwiki,cmssusytwiki}.  However, direct searches must make assumptions about the decay of the gluino.  Indirect searches are most useful when the gluino decays in a way that is difficult or impossible to identify.  There are many possible ways in which a low-mass gluino could evade direct searches - see Ref.~\cite{Evans:2018scg} for a recent example suggesting that a 50 GeV gluino may be under-constrained.  More generally, indirect searches play an important role when new colored particles of any kind are not possible to identify through direct means.

The goal of this paper is to reinterpret the recent ATLAS measurement of event shapes using jets.  Transverse energy-energy correlations (TEEC) (Sec.~\ref{sec:teecs}) are measured as a function of jet $p_\mathrm{T}$, which are sensitive to the running of $\alpha_s$.  While the aim of this analysis is similar to Ref.~\cite{spannowsky}, there are several key distinctions.   First, instead of reinterpreting $\alpha_s$ measurements, a global fit is performed directly to the measured data.  This allows a probe of the entire shape of the TEEC distribution\footnote{Since both this analysis and Ref.~\cite{spannowsky} ignore real emissions, $\alpha_s$ is actually a sufficient statistic of the data.  In general, if adding new physics changed more than just the scale dependence of $\alpha_s(Q)$, then it would be preferable to use the full distribution to avoid bias.}.  Furthermore, the entire experimental covariance matrix is used when performing the fit.  The correlations across $p_\mathrm{T}$ is particularly important and can actually result in worse limits than if the uncertainties were assumed independent.  Finally, the analysis presented in this paper also includes a detailed study of theoretical systematic uncertainties.  Such uncertainties are not small and have a significant impact on the results.  The theoretical predictions and statistical analysis are described in detail in Sections~\ref{sec:theory} and~\ref{sec:statanalysis}, respectively, before presenting the results in Sec.~\ref{sec:results}.  The paper ends in Sec.~\ref{sec:conclusions} with conclusions and prospects for the future.



\section{Transverse energy-energy correlations}
\label{sec:teecs}
The energy-energy correlation function (EEC)~\cite{basham1, basham2}, defined as the energy-weighted azimuthal differences between pairs of hadrons, has been widely used in $e^{+} e^{-}$ colliders~\cite{alephEEC, delphiEEC} as a precision test of QCD and as a means to determine the strong coupling constant. This observable was later adapted to hadron-hadron colliders by using its projection on the transverse plane (TEEC)~\cite{ali1}, defined as a function of the angle $\phi$ between two jets in a given event
\begin{equation}
\frac{1}{\sigma}\frac{\mathrm{d}\Sigma}{\mathrm{d}\cos\phi} \equiv \frac{1}{\sigma}\sum_{ij}\int\frac{\mathrm{d}\sigma}{\mathrm{d}x_{\mathrm{T}i} \mathrm{d}x_{\mathrm{T}j} \mathrm{d}\cos\phi}x_{\mathrm{T}i}x_{\mathrm{T}j} \mathrm{d}x_{\mathrm{T}i} \mathrm{d}x_{\mathrm{T}j},
\label{eq:teecDef}
\end{equation}
where $\sigma$ is the total inclusive two-jet cross section and $x_{\mathrm{T}i}$ is the fraction of energy carried by jet $i$ with respect to the total:
\begin{equation}
x_{\mathrm{T}i} = \frac{E_{\mathrm{T}i}}{\sum_k E_{\mathrm{T}k}}.
\label{eq:xtDef}
\end{equation}

The TEEC is useful for measuring the strong coupling constant because its distribution is proportional to $\alpha_s$ and yet as a ratio observable, multiple sources of theoretical and experimental uncertainty cancel.  Next-to-leading (NLO) corrections to the TEEC function have been calculated for $pp$ collisions in Ref.~\cite{ali2}. Most recently, the ATLAS Collaboration has published measurements of the TEEC and its asymmetry 
\begin{equation}
\frac{1}{\sigma}\frac{\mathrm{d}\Sigma^{\rm asym}}{\mathrm{d}\cos\phi}\equiv
\left.\frac{1}{\sigma}\frac{\mathrm{d}\Sigma}{\mathrm{d}\cos\phi}\right|_{\phi} -
\left.\frac{1}{\sigma}\frac{\mathrm{d}\Sigma}{\mathrm{d}\cos\phi}\right|_{\pi -\phi},
\label{eq:ateecDef}
\end{equation}
at $\sqrt{s}=7$ and $8$ TeV~\cite{atlas1, atlas2} up to $Q\sim 800$~GeV, from which the strong coupling constant was extracted and its scale evolution was precisely studied. These data provides not only a precise measurement of the angular correlations in multijet final states, but also a handle with which to study possible contributions of new physics to the virtual corrections to the gluon propagator, which would modify the QCD $\beta$ function and therefore the solution of the renormalization group equation.

\section{Theoretical predictions}
\label{sec:theory}

The one-loop virtual corrections to the quark and gluon propagators in perturbative QCD are parameterised by means of the renormalization group equation (RGE)
\begin{equation}
\frac{\partial\alpha_s}{\partial\log Q^2} = \beta(\alpha_s) = -\alpha_s^2(\beta_0+\beta_1\alpha_s+\mathcal{O}(\alpha_s^2)),
\label{eq:renGroup}
\end{equation}
where the coefficients $\beta_0$ and $\beta_1$ are given by
\begin{equation}
\beta_0 = \frac{1}{4\pi}\left(11-\frac{2}{3} n_f\right); \ \ \ \beta_1 = \frac{1}{(4\pi)^2}\left(102-\frac{38}{3}n_f\right),
\label{eq:betaQCD}
\end{equation}
and $n_f$ is the number of active quark flavours at the scale $Q$. The presence of additional fermions entering the loops would modify these coefficients as \cite{spannowsky}
\begin{eqnarray}\label{eq:betaMod1}
\beta_0 = \frac{1}{4\pi}\left(11-\frac{2}{3} n_f- \frac{4}{3}n_X T_X \right)\\
\label{eq:betaMod2}
\beta_1 = \frac{1}{(4\pi)^2}\left[102-\frac{38}{3}n_f -20 n_X T_X \left(1+\frac{C_X}{5}\right)\right].
\end{eqnarray} 
The new terms in Eq. \ref{eq:betaMod1} and \ref{eq:betaMod2} include the number of new fermions, $n_X$, transforming under a given representation of $SU(3)$ parameterised by the group factor $T_X$, as well as the Casimir $C_X$. Examples of such fermions are Dirac triplets, octets, sextets and decuplets transforming under representations of dimension 3, 8, 6 and 10, respectively. For these particular models, the values of $T_X$ and $C_X$ are given in Table~\ref{tab:models}.  The leading order modification can be parameterized by the mass of the new particles and $n_\mathrm{eff}=2\sum n_X T_X$.  For example, the addition of a gluino would result in $n_\mathrm{eff}=3$ and the entire Minimal Supersymmetric Standard Model (MSSM) would have $n_\mathrm{eff}=6$.  In principle, $n_\mathrm{eff}$ need not be an integer, as might be the case if there is a strongly coupled dark sector that communicates with the SM QCD as in Ref.~\cite{Cohen:2017pzm}.  Only integer values are used for predictions in the subsequent sections.

\begin{table}[H]
\begin{center}
\begin{tabular}{c|cccc}
\T\B \bf{Model} & Triplet & Octet & Sextet & Decuplet\\
\hline
\T\B \bf{$T_X$} & 1/2 & 3 & 5/2 & 15/2\\
\T\B \bf{$C_X$} & 4/3 & 3 & 10/3 & 6
\end{tabular}
\end{center}
\caption{Example values of $T_X$ and $C_X$ for Dirac triplets, octets, sextets and decuplets.}
\label{tab:models}
\end{table}

\begin{figure}[H]
\centering
\includegraphics[width=12.cm,height=8.5cm]{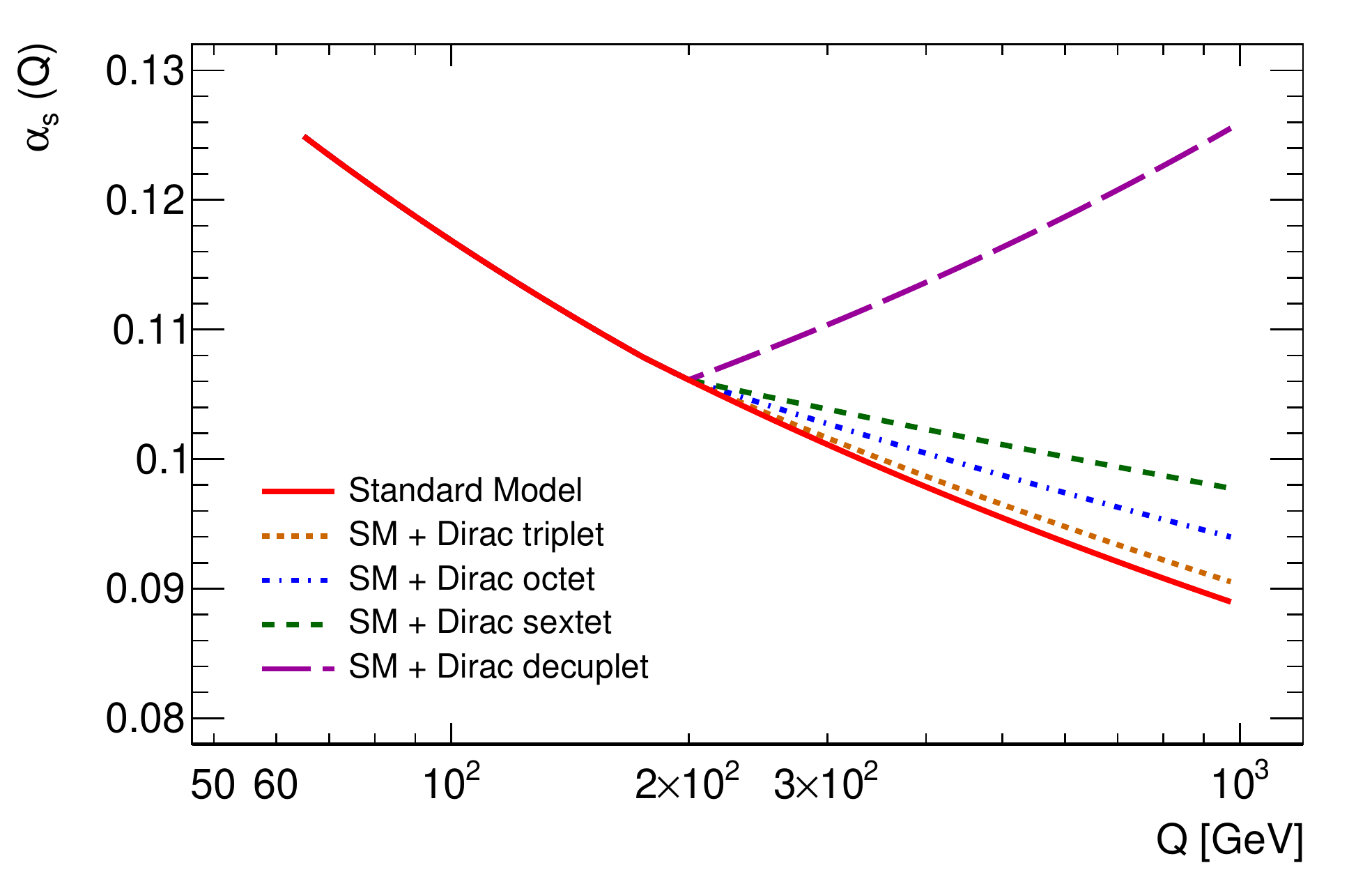}
\caption{The next-to-leading order solutions to the renormalization group equation including a new fermion with mass $m_X = 200$ GeV transforming under representations of dimension 3, 8, 6 and 10, respectively.}
\label{fig:runAs}
\end{figure}

The evolution of the strong coupling constant for the four particular models listed in Table~\ref{tab:models} is shown in Fig.~\ref{fig:runAs}. As noted in Ref. \cite{spannowsky}, for some values of $n_{\rm{eff}}$, asymptotic freedom may be lost.

At next-to-leading order, the transverse energy-energy correlation function can be expressed as a second-order polynomial in $\alpha_s(Q)$, i.e. \cite{ali2}
\begin{equation}
\frac{1}{\sigma}\frac{\mathrm{d}\Sigma}{\mathrm{d}\cos\phi} \propto \frac{\alpha_s(Q)}{\pi}F(\phi)\left[1+\frac{\alpha_s(Q)}{\pi}G(\phi)\right],
\label{eq:teecNLO}
\end{equation}
where $F(\phi)$ and $G(\phi)$ are functions of the azimuth to be determined in the perturbative calculation; and $\alpha_s(Q)$ is the solution to Eq. \ref{eq:renGroup}, which at NLO is given by
\begin{equation}
\alpha_s(Q) = \frac{1}{\beta_0\log z}
\left[1- \frac{\beta_1}{\beta_0^2}
\frac{\log\left(\log z \right)}{\log z}\right]; \ \ z = \frac{Q^2}{\Lambda_{\mathrm{QCD}}^2}.
\label{eq:2loop}
\end{equation}

\begin{figure}[H]
\centering
\includegraphics[width=6.8cm,height=8.cm]{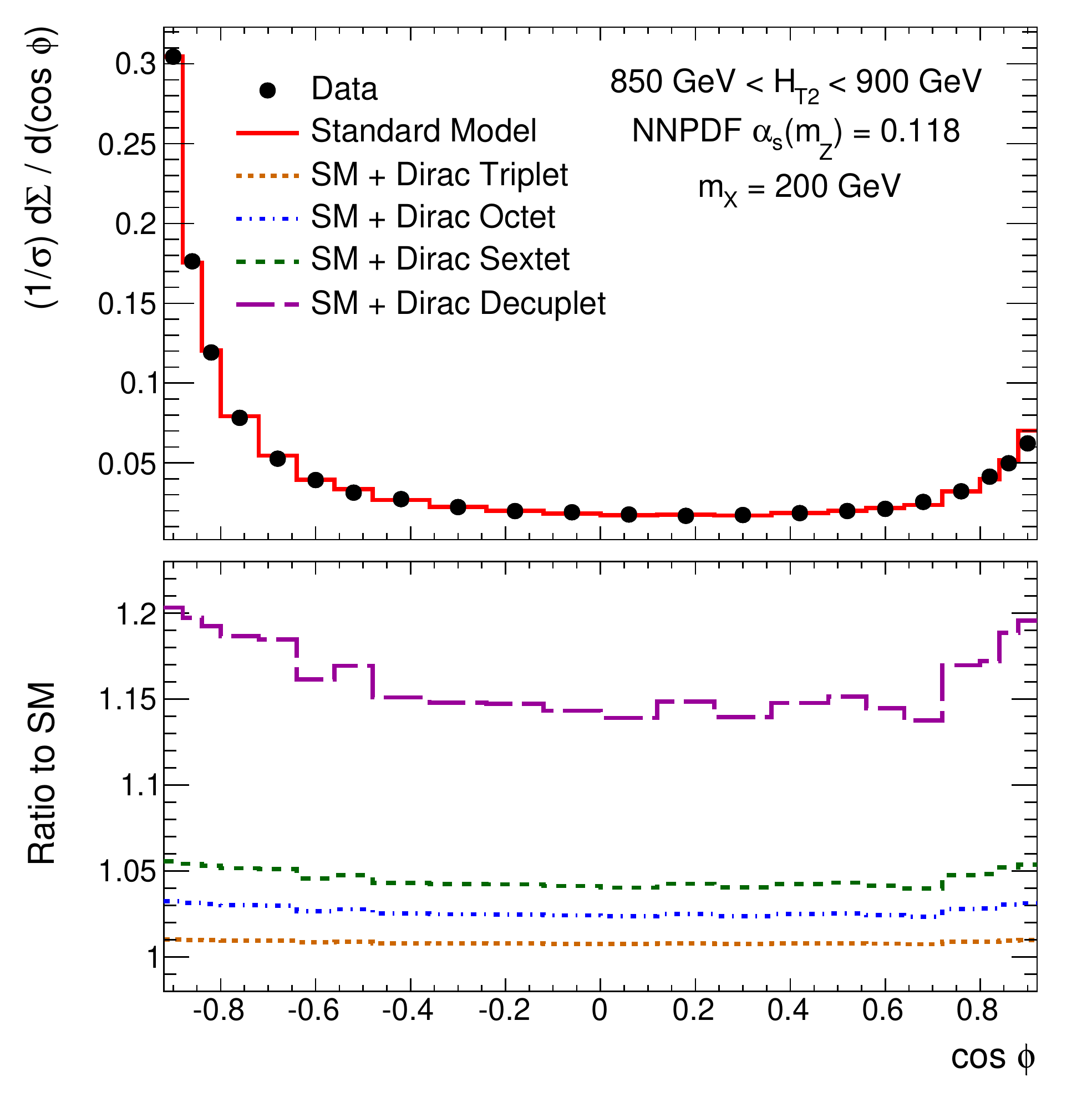}
\includegraphics[width=6.8cm,height=8.cm]{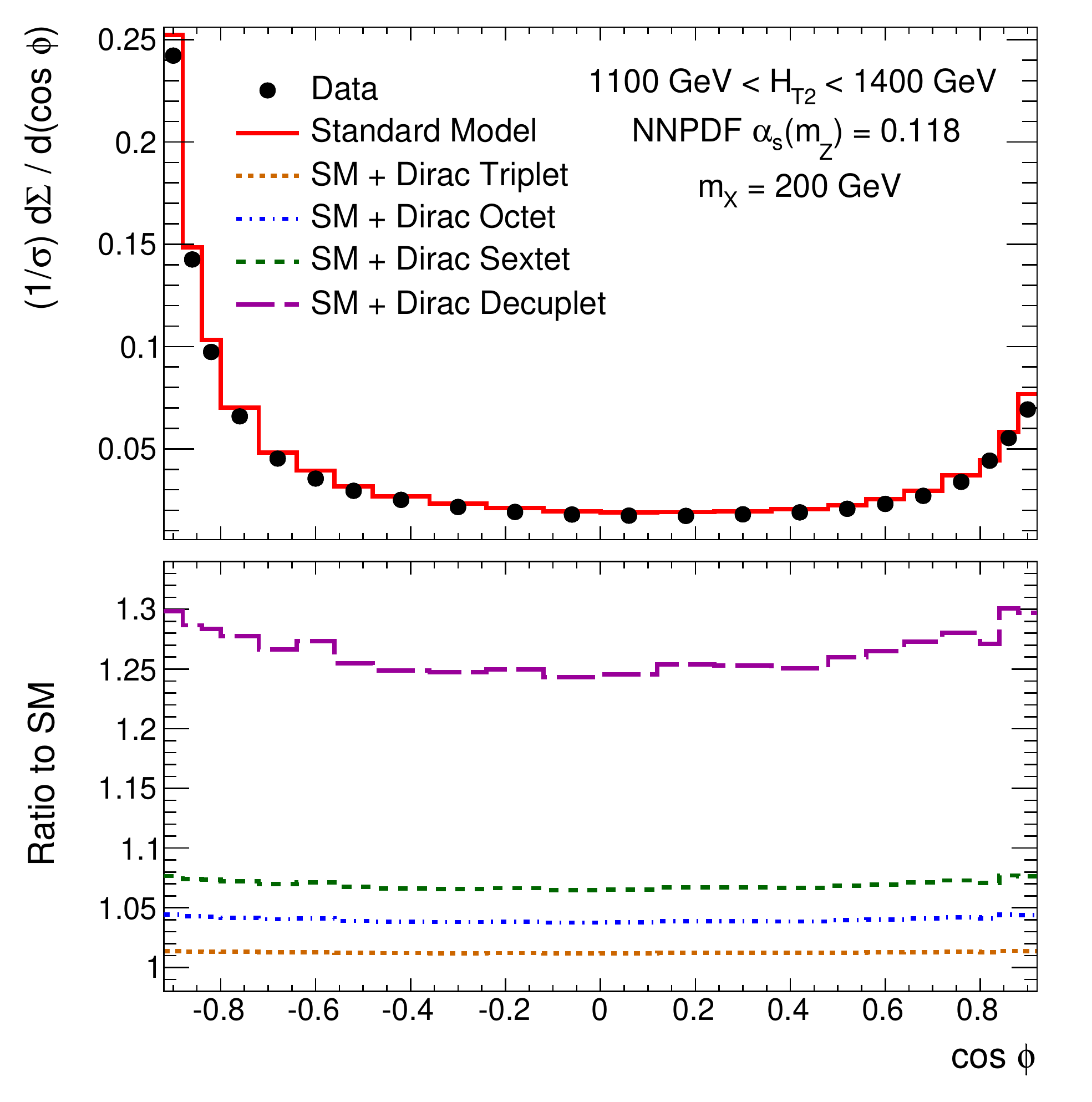}

\includegraphics[width=6.8cm,height=8.cm]{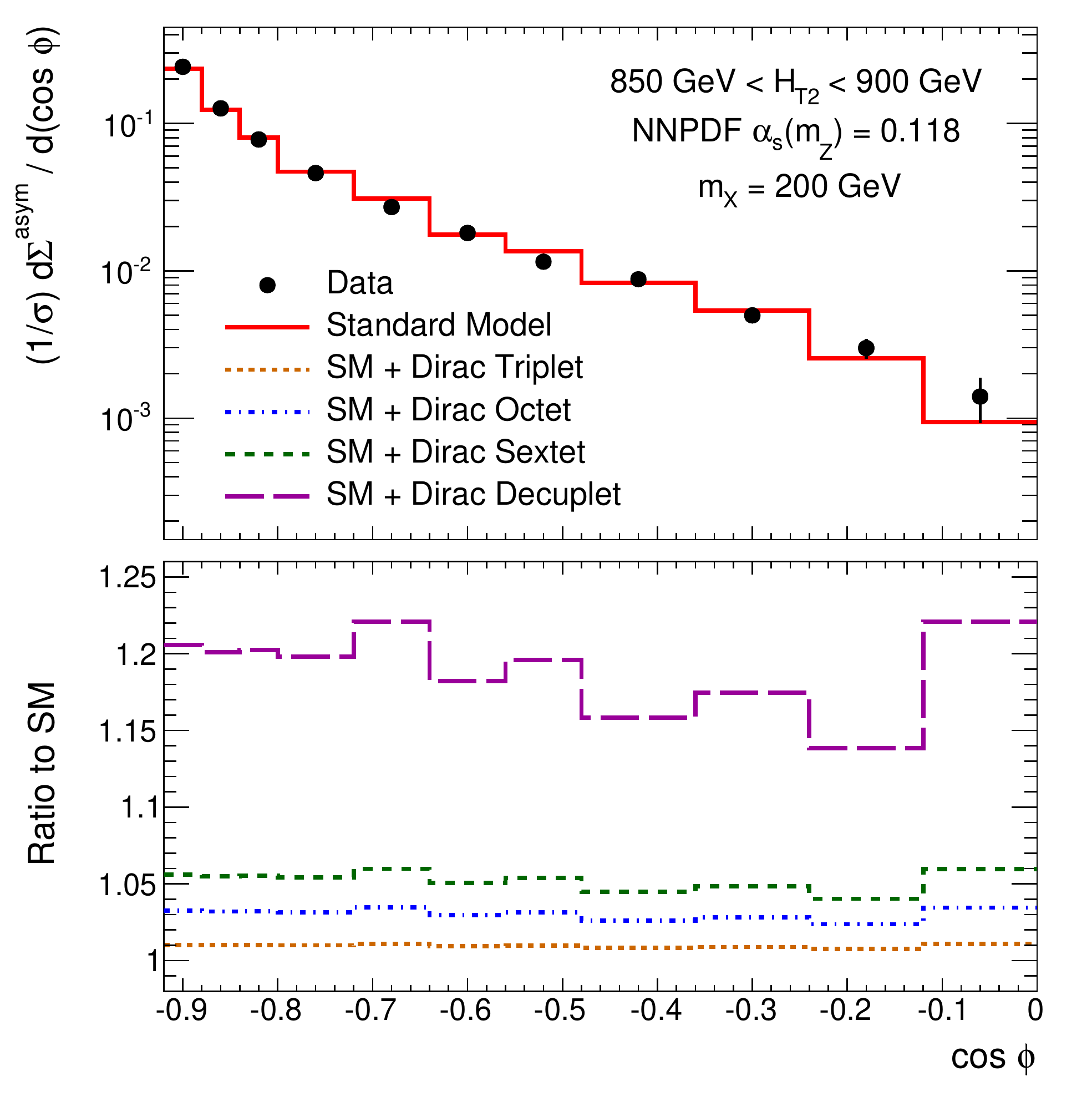}
\includegraphics[width=6.8cm,height=8.cm]{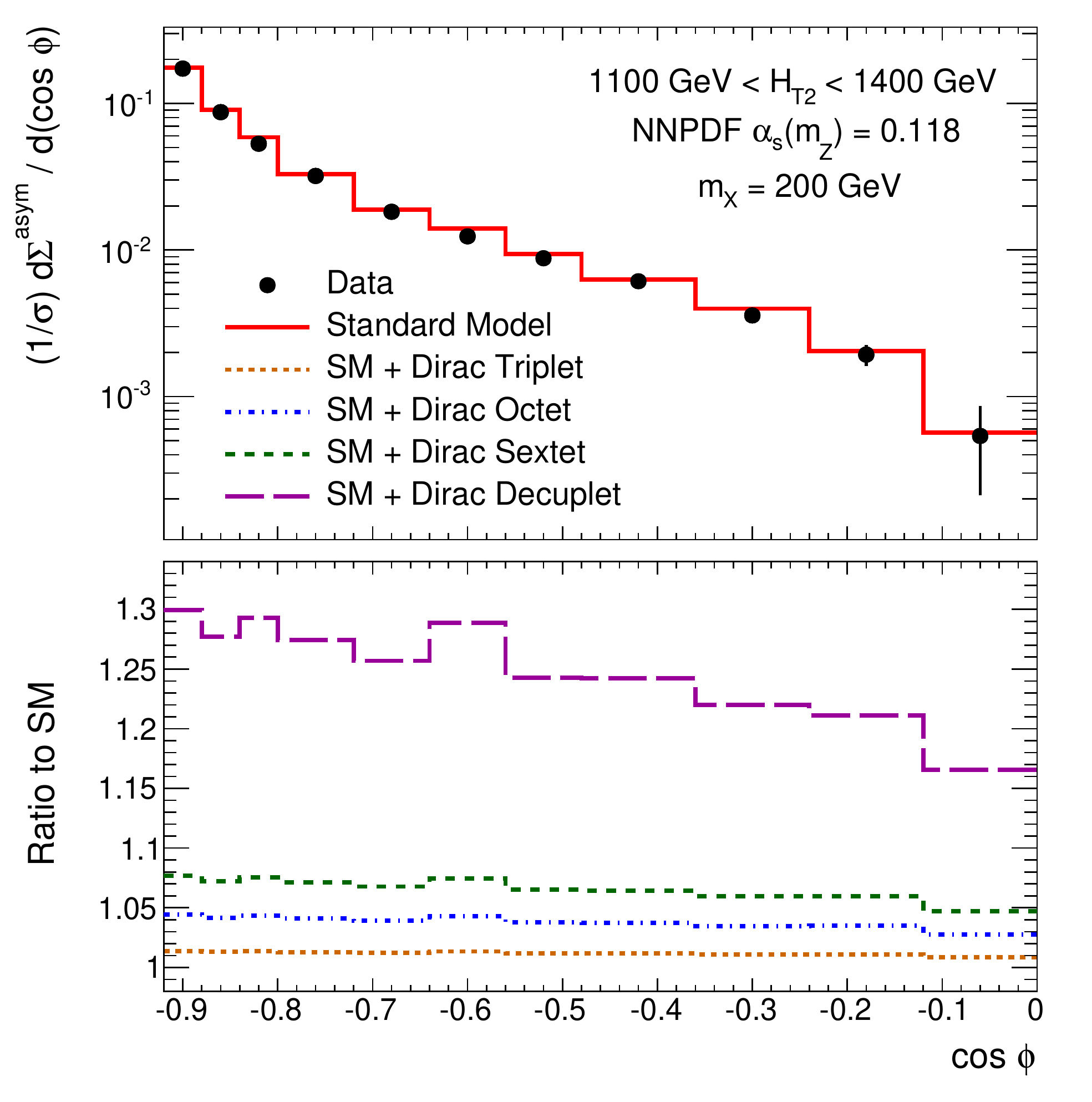}
\caption{Theoretical predictions for the TEEC (top) and ATEEC functions (bottom) for two sample intervals of $H_{\mathrm{T}2}$ provided by the ATLAS measurement~\cite{hepdata,atlas2}. The Standard Model prediction is shown, together with the data, on the top panel for each subfigure. The bottom panels display the ratio of the theoretical predictions in four sample BSM models with respect to the Standard Model prediction, showing a clear increase for each bin of the distribution.}
\label{fig:teecBSM}
\end{figure}

With these ingredients, one can obtain the theoretical predictions for the TEEC functions using NLOJet++~\cite{nagy1, nagy2}, together with the NNPDF 3.0 parton distribution functions~\cite{nnpdf}  In addition to the truncation in the fixed-order perturbative series, the resulting predictions for the TEEC functions involve three additional approximations.  The first approximation is independent of BSM: we avoid from regions of phase space with significant collinear enhancement since we do not include higher order resummation in the calculation.  In practice, this is accomplished by restricting $\cos\phi$ to be away from $\pm 1$. Next, we neglect the impact of new fermions on the proton parton distribution functions (PDFs).  This is justified because the TEEC is a ratio of 3-jet to 2-jet cross-sections and so the effects of PDF variations largely cancel~\cite{ali2}.  This is further supported by the fact that the theoretical uncertainties due to the PDF were shown to be negligible in Ref.~\cite{atlas1, atlas2}.  A more detailed analysis in Ref.~\cite{spannowsky} also found that the contribution from PDF variations was negligible for ratio observables.  The third approximation is that we neglect real emissions of new fermions.  When they are abundantly produced and readily detected, direct searches for new colored particles are more effective than indirect searches.  Therefore, it is safe to assume that the direct production of such particles is suppressed.  This is also supported by the fact that the event selections used in Ref.~\cite{atlas1, atlas2} are very inclusive and not highly specialized as in the direct searches.  Additionally, we have explicitly checked the gluino models from Ref.~\cite{kaplan} using MG5\_aMC@NLO 2.6.0~\cite{Alwall:2014hca} with the RPV model~\cite{Fuks:2012im} and find that the contribution to the event selection for Ref.~\cite{atlas1, atlas2} is negligible.
Figure~\ref{fig:teecBSM} shows the predictions for the TEEC functions described above, under various (B)SM scenarios and compared with ATLAS data from Ref.~\cite{hepdata,atlas2}.  The events are required to have at least two jets with $p_{\mathrm{T}}>100$ GeV and $|\eta|<2.5$.  The scalar sum of the transverse momenta of the leading two jets ($H_{\mathrm{T2}}$) must be greater than 800 GeV. Following the procedure in Ref.~\cite{atlas2}, the renormalization scale, at which $\alpha_s(Q)$ is evaluated, is taken to be $\mu_R = H_{\mathrm{T2}}/2$, while the factorization scale is set to $\mu_F = \mu_R/2$.  In Fig.~\ref{fig:teecBSM}, the new fermion has a mass $m_X = 200$ GeV transforming under the four different representations described above (triplet, octet, sextet and decuplet). From this figure, the effect of the running of $\alpha_s$ on the predicted distributions is very clear, showing the increase of both the TEEC and ATEEC distributions caused by the modification of the $\beta$ function.

\newpage

\newpage
\section{Statistical analysis}
\label{sec:statanalysis}

\begin{figure}[H]
\centering
\includegraphics[width=0.5\textwidth]{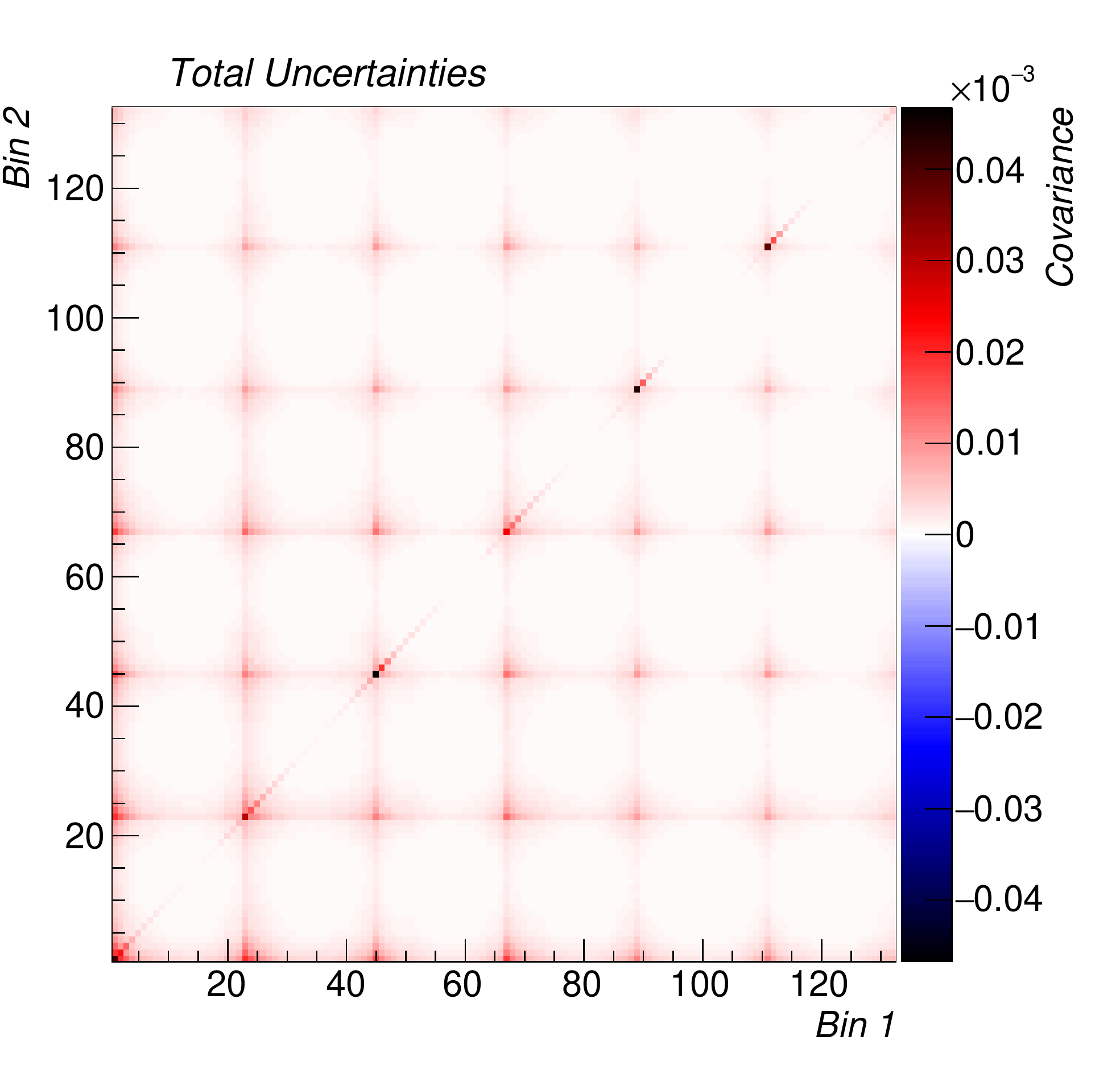}\includegraphics[width=0.5\textwidth]{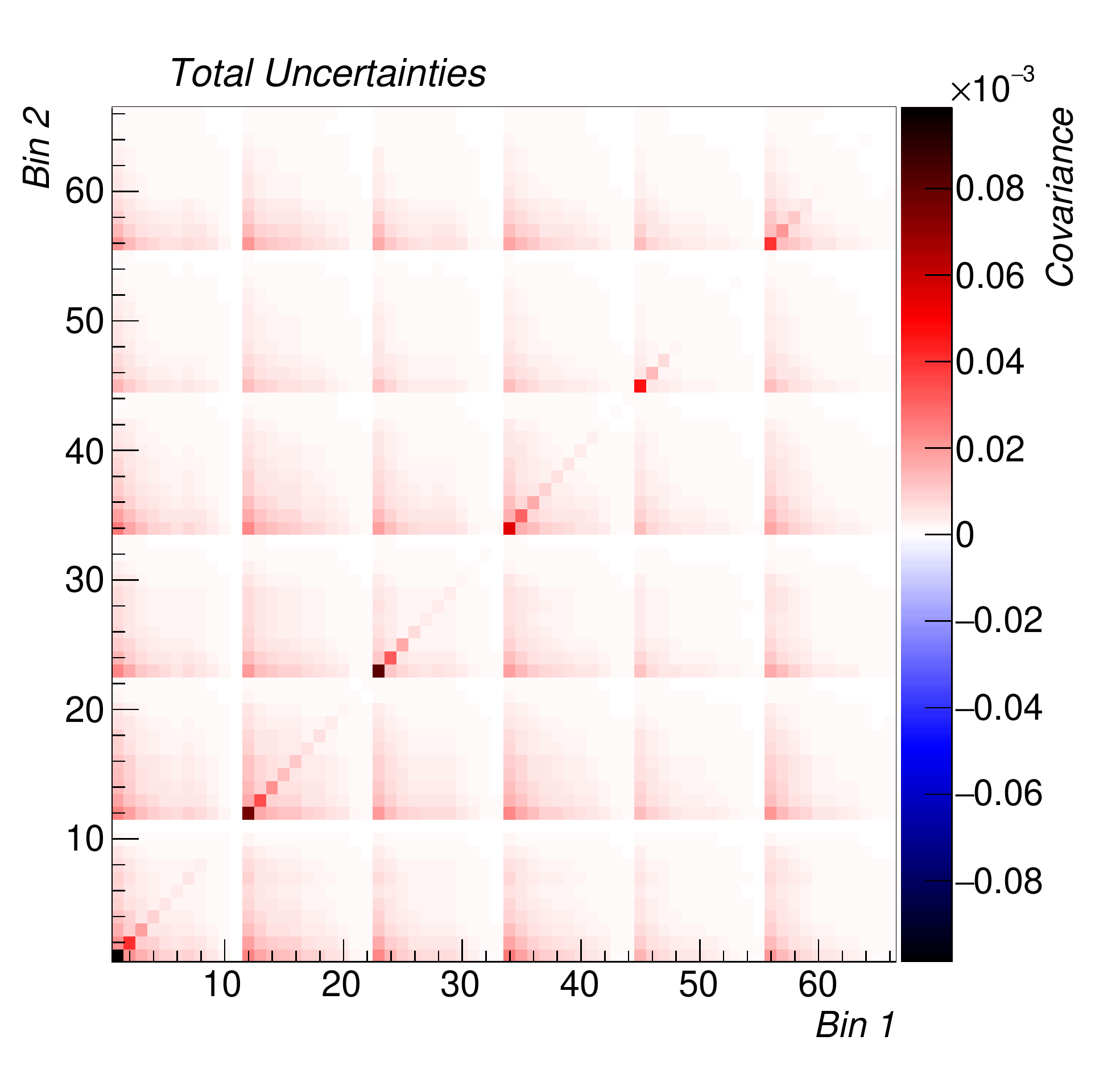}
\caption{A visualization of $\Sigma$  for the TEEC (left) and for the ATEEC (right).}
\label{fig:matrices}
\end{figure}

Pseudo-experiments are constructed in order to assess the $p$-value for a particular BSM hypothesis.  The generative model for the result of a given pseudo-experiment $X$ is given by

\begin{equation}
\label{eq:generative}
X=X_{\mathrm{stat}}+\sum_{i=1}^{n_{\mathrm{sys}}}\left(\sigma_i^+X_i\mathbb{I}[X_i>0]+\sigma_i^-X_i\mathbb{I}[X_i\leq 0]\right),
\end{equation}

\noindent where $X_{\mathrm{stat}}$ is a multidimensional Gaussian random variable with mean values given by a prediction and the covariance matrix is from the statistical uncertainty.  The statistical uncertainty at detector-level is a product of Poisson random variables, but after unfolding, there are correlations between bins.  Due to the large number of events, these fluctuations are well-modeled as normally distributed.  There are $n_{\mathrm{sys}}$ total systematic uncertainties (see Ref.~\cite{hepdata,atlas2} for details) and $\sigma_i^\pm\in\mathbb{R}^{n_\mathrm{bins}}$ for $n_\mathrm{bins}$ total measurement bins; $\sigma$ corresponds to the asymmetric $\pm$ uncertainty on each bin.  In order to remove bins that are not well-descibed by fixed-order QCD, the first (and last) three bins from each $H_\text{T2}$ bin are removed from the ATEEC (TEEC) distribution.  The random variable $X$ is distributed as a standard normal random variable.  This setup treats the systematic uncertainties as fully correlated across all bins.  These correlations are expected to be accurate except for the case of modeling uncertainties, as they are constructed from comparing two models of fragmentation.  Therefore, the modeling uncertainty is treated as uncorrelated across bins.  According to the Neyman-Pearson lemma~\cite{Neyman289}, the most powerful statistical test uses the likelihood ratio statistic.  It is possible to analytically compute the likelihood from Eq.~\ref{eq:generative}, though it is very cumbersome due to the asymmetric systematic uncertainties.  Numerically, the largest $\sigma_{i}^+$ are within 10\% of $\sigma_{i}^-$ and so for a test statistic, a symmetrized likelihood based on $\sigma_i=\frac{1}{2}(\sigma_i^++\sigma_i^-)$ is used instead.  In that case, the log likelihood (up to a constant) can be written as

\begin{align}
\label{sec:teststat}
\log(p(X|\theta))=-\frac{1}{2}(X-\theta)^T\Sigma^{-1}(X-\theta),
\end{align}

\noindent where $\theta\in\mathbb{R}^{n_\mathrm{bins}}$ is the predicted distribution and $\Sigma=\Sigma_\text{stat}+\sum_{i=1}^{n_{\mathrm{sys}}}\Sigma_i$.  The systematic uncertainty covariance matrix $(\Sigma_{i})_{jk}=\sigma_{ij}\sigma_{ik}$ where $\sigma_{ij}$ is the (signed) uncertainty for systematic uncertainty $i$ in bin $j$.  Note that $\Sigma_i$ is singular (has rank 1) and so only the sum of all covariance matrices results in a well-behaved multi-dimensional probability distribution.  The modeling uncertainty covariance matrix only includes the diagonal entries.  Equation~\ref{sec:teststat} is used as the test statistic for performing hypothesis tests with the data.  Figure~\ref{fig:matrices} shows the full covariance matrices for the TEEC and ATEEC cases.  The repeated structures indicate the $H_\text{T2}$ binning and are due to strong correlations between these bins.  The diagonal component is due mostly to the statistical uncertainty and the modeling uncertainty.

\begin{figure}[H]
\centering
\includegraphics[width=0.6\textwidth]{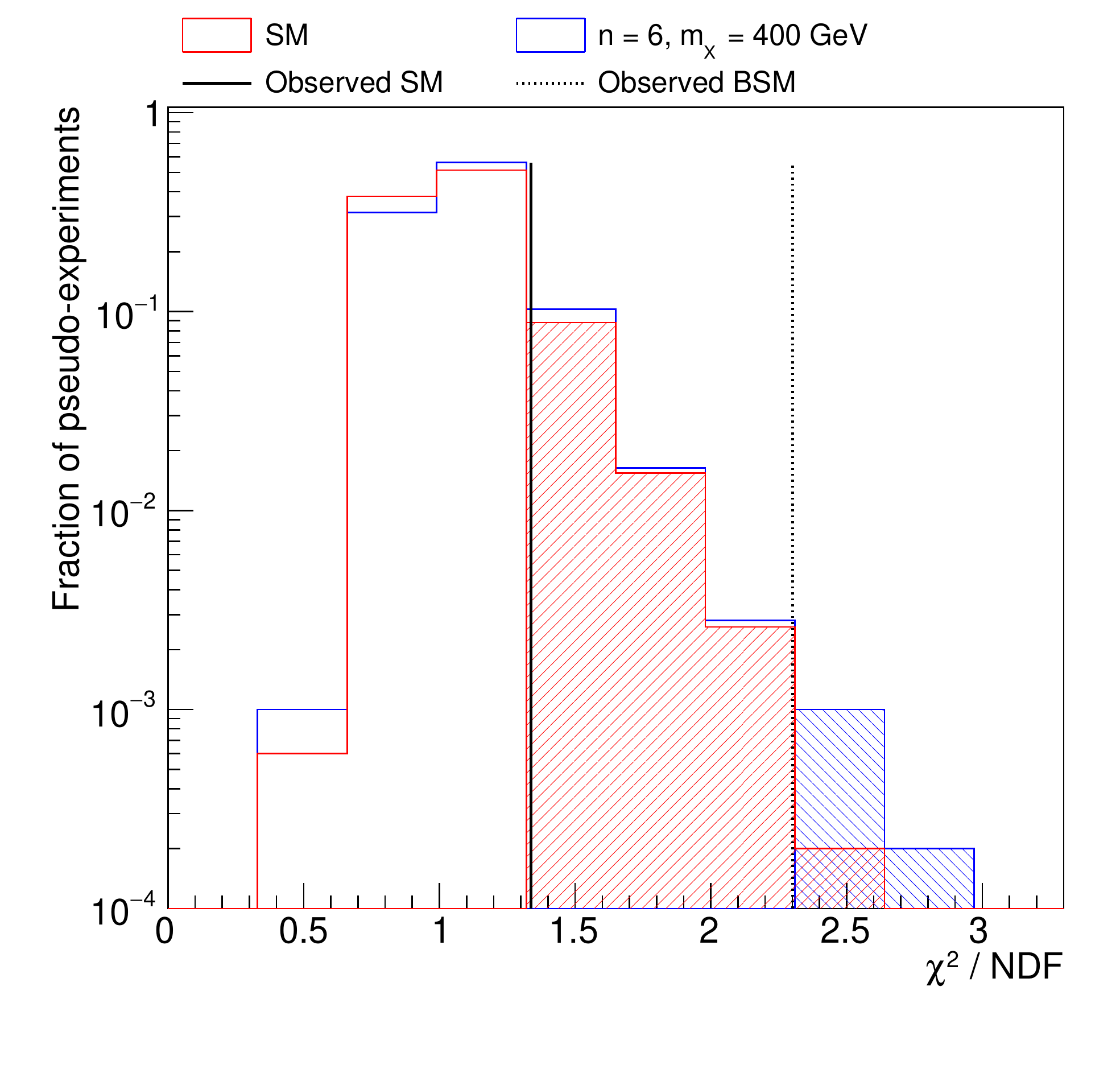}
\caption{The distribution of $\chi^2$ under the SM-only hypothesis (red) and a particular BSM scenario (blue) with $n_{\mathrm{eff}}=6$ and $m_\chi=300$ GeV.  The distributions are computed using pseudo-experiments, as described in the text.  Vertical lines indicate the $\chi^2$ values corresponding to the ATLAS measurement.  }
\label{fig:chi2distributions}
\end{figure}

Figure~\ref{fig:chi2distributions} shows the distribution of the $\chi^2$ for many pseudo-experiments for the SM (red) and $n_{\mathrm{eff}}=6$, $m_X=400$ GeV (blue).  The expected $\chi^2$ distribution is nearly the same for the SM and BSM case, though the observed values are very different - the data are much less likely under the BSM hypothesis.  In particular, the SM is consistent with the prediction at the $1.3\sigma$ level while the tail probability for the BSM model shown is about $3\sigma$.

\clearpage

\section{Results}
\label{sec:results}

\begin{figure}[H]
\centering
\includegraphics[width=0.5\textwidth]{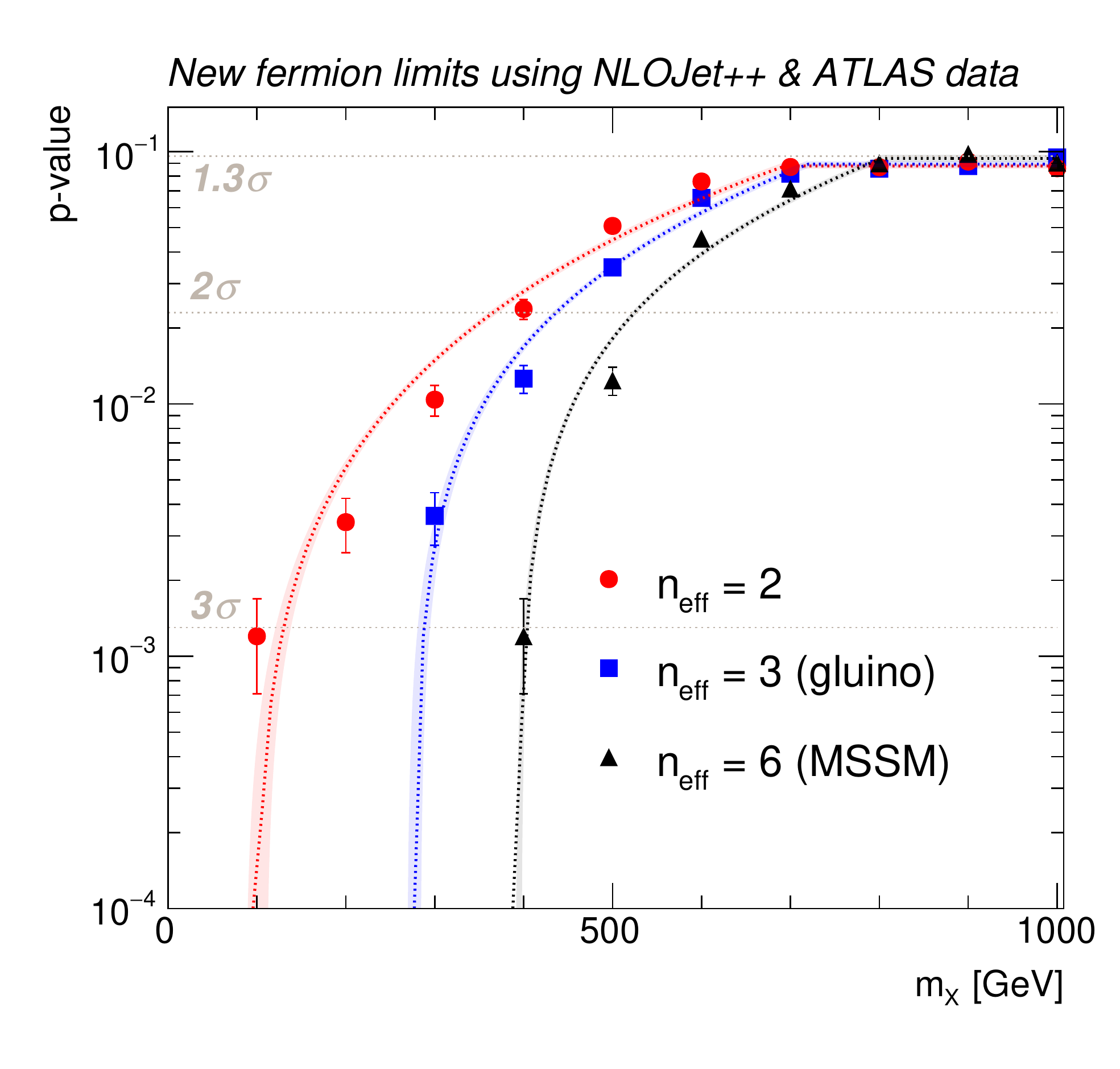}\includegraphics[width=0.5\textwidth]{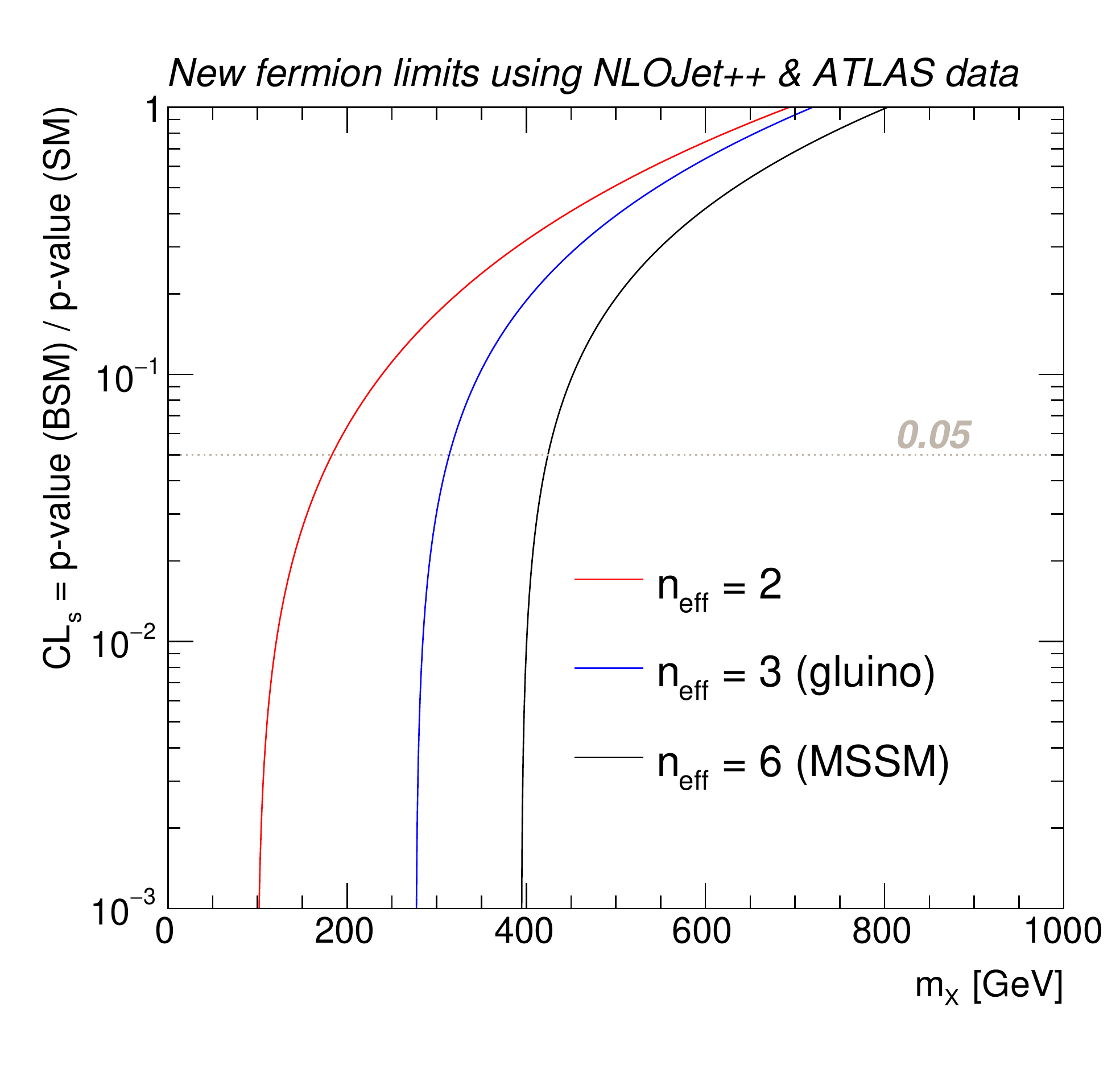}
\caption{The computed $p$-values (left) and fitted CL$_\mathrm{s}$ (right) for three BSM models corresponding to $n_{\mathrm{eff}}=2,3$, and $6$ as a function of the new particle mass. In each case, the $p$-value asymptotes to about $1.5\sigma$, which is the SM value. Fitted quadratic curves asymptoting to a flat line are shown to guide the eye.}
\label{fig:pvalues}
\end{figure}

For three values of $n_{\mathrm{eff}}$, Fig.~\ref{fig:pvalues} shows the distribution of the $p$-value for the data under the BSM hypothesis specified by the mass value.  For sufficiently high values of $m_X$, the impact of the BSM is negligible and the predictions all converge to the SM.  As in Fig.~\ref{fig:chi2distributions}, the SM prediction is consistent with the data at the 1.3$\sigma$ level.  Low masses and higher dimensional QCD representations result in larger differences with the SM and also with the data.  The $n_{\mathrm{eff}}=2$, $n_{\mathrm{eff}}=3$ and $n_{\mathrm{eff}}=6$ cases crosses $2\sigma$ around 350 GeV, 400 GeV, and 500 GeV, respectively.

The usual procedure for setting limits at the LHC uses the CL$_\mathrm{s}$ approach~\cite{0954-3899-28-10-313}, which penalizes discrepancies with the SM and overall lack of sensitivity to BSM by taking a ratio of $p$-values.  By definition, the CL$_\mathrm{s}$ value is the ratio of the $p$-value under the BSM hypothesis divided by the $p$-value under the SM-only hypothesis.  The right-hand side of Fig.~\ref{fig:pvalues} is the same as the left-hand side, only each value is divided by the SM $p$-value.  By construction, this CL$_\mathrm{s}$ value approaches unity as $m_X\rightarrow\infty$.  A model is declared `excluded' when this value crosses below $0.05$.  The corresponding limits for $n_{\mathrm{eff}}=2$, $n_{\mathrm{eff}}=3$ and $n_{\mathrm{eff}}=6$ are about 200 GeV, 300 GeV, and 400 GeV, respectively.

\begin{figure}[H]
\centering
\includegraphics[width=0.5\textwidth]{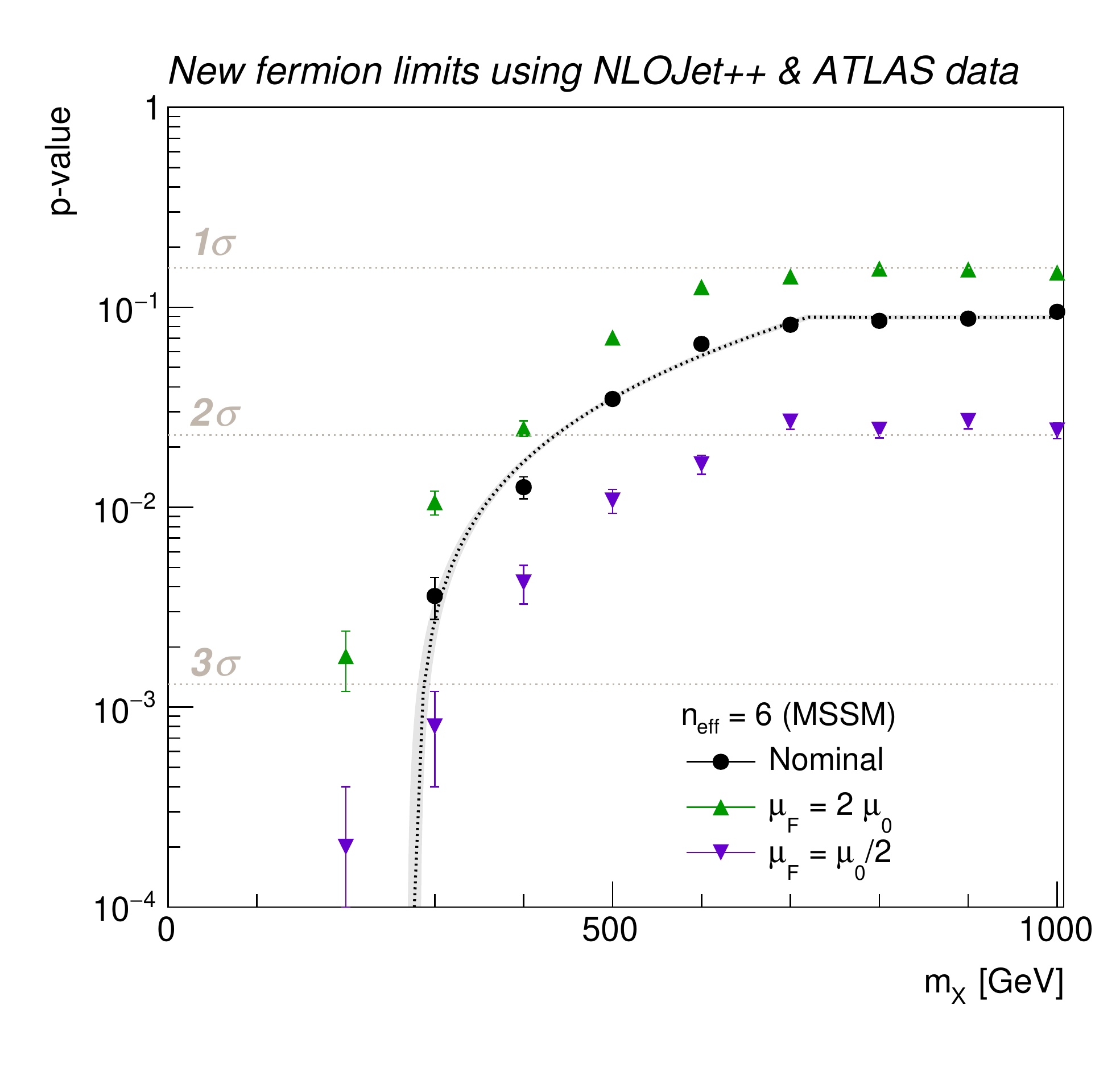}\includegraphics[width=0.5\textwidth]{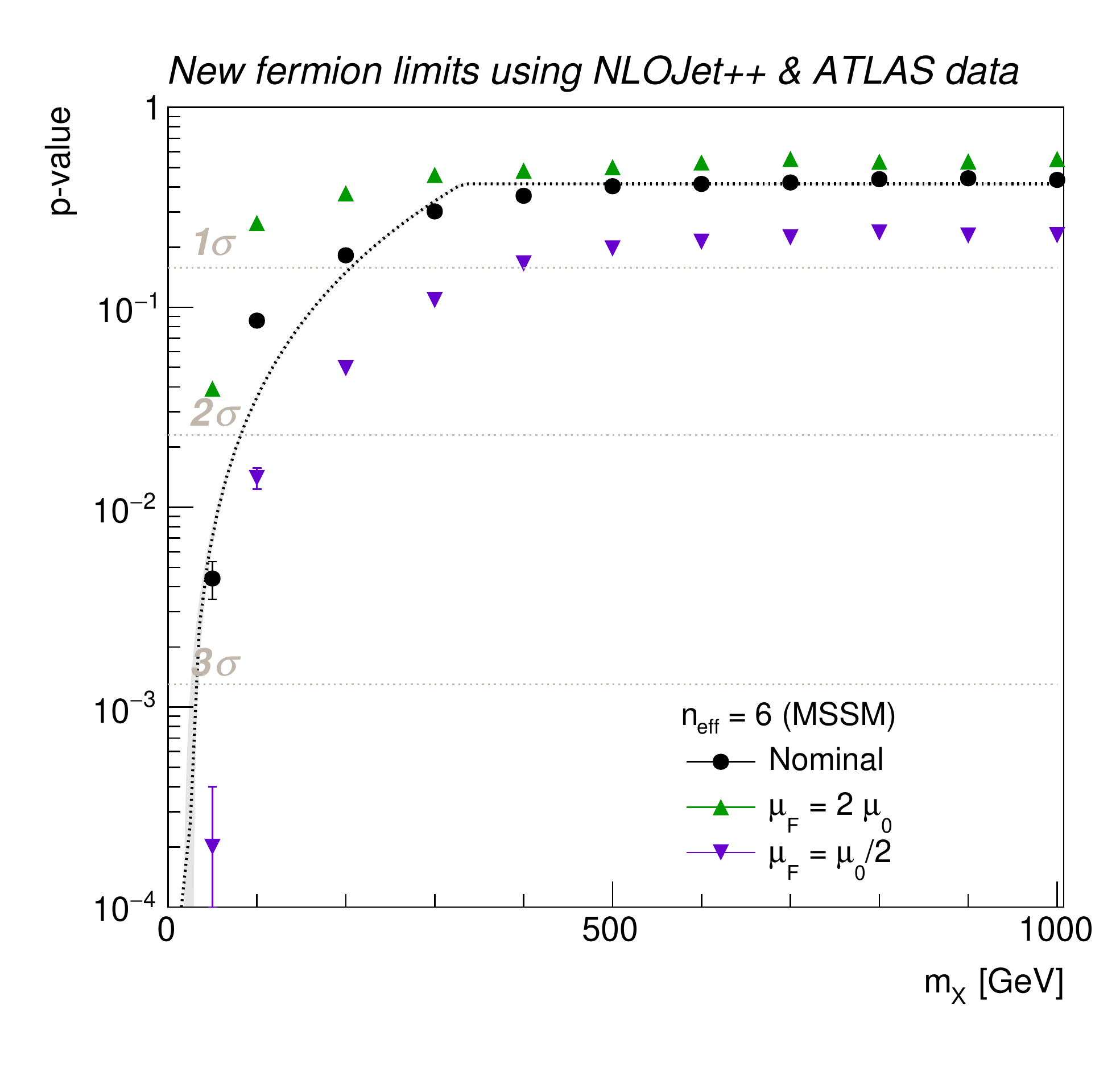}
\caption{The same $p$-value as shown in Fig.~\ref{fig:pvalues}, but now only for $n_{\mathrm{eff}}=6$.  Additional colored points indicate the values when the factorization scale is varied up and down by factors of $2$.  The left plot is for the TEEC and the right plot is for the ATEEC.}
\label{fig:uncerts}
\end{figure}

All of the results presented thus far have ignored theoretical uncertainties, except PDF and fragmentation uncertainties.  The results in Ref.~\cite{atlas2} showed that the theoretical uncertainty from varying the factorization and renormalization scales by factors of $1/2$ and $2$ can significantly exceed the experimental uncertainty. Given that the renormalization scale $\mu_R$ is the argument of $\alpha_s(Q)$ in the fixed order calculation, it is expected that varying it by a factor of 2 up and down will impact the exclusion limit roughly by a factor of 2. The factorization scale variations are shown in Fig.~\ref{fig:uncerts}. However, given the degree of arbitrariness in the choice of the QCD scales $(\mu_R, \mu_F)$, these variations are shown only for illustrative purposes, and should necessarily not be taken as theoretical uncertainties on the exclusion limits.\\
Uncertainties due to the precision in $\alpha_s(m_Z)$ have also been evaluated by varying its value by 1\%~\cite{nnpdf}. These variations have a negligible impact on the exclusion limits.

\begin{figure}[H]
\centering
\includegraphics[width=0.8\textwidth]{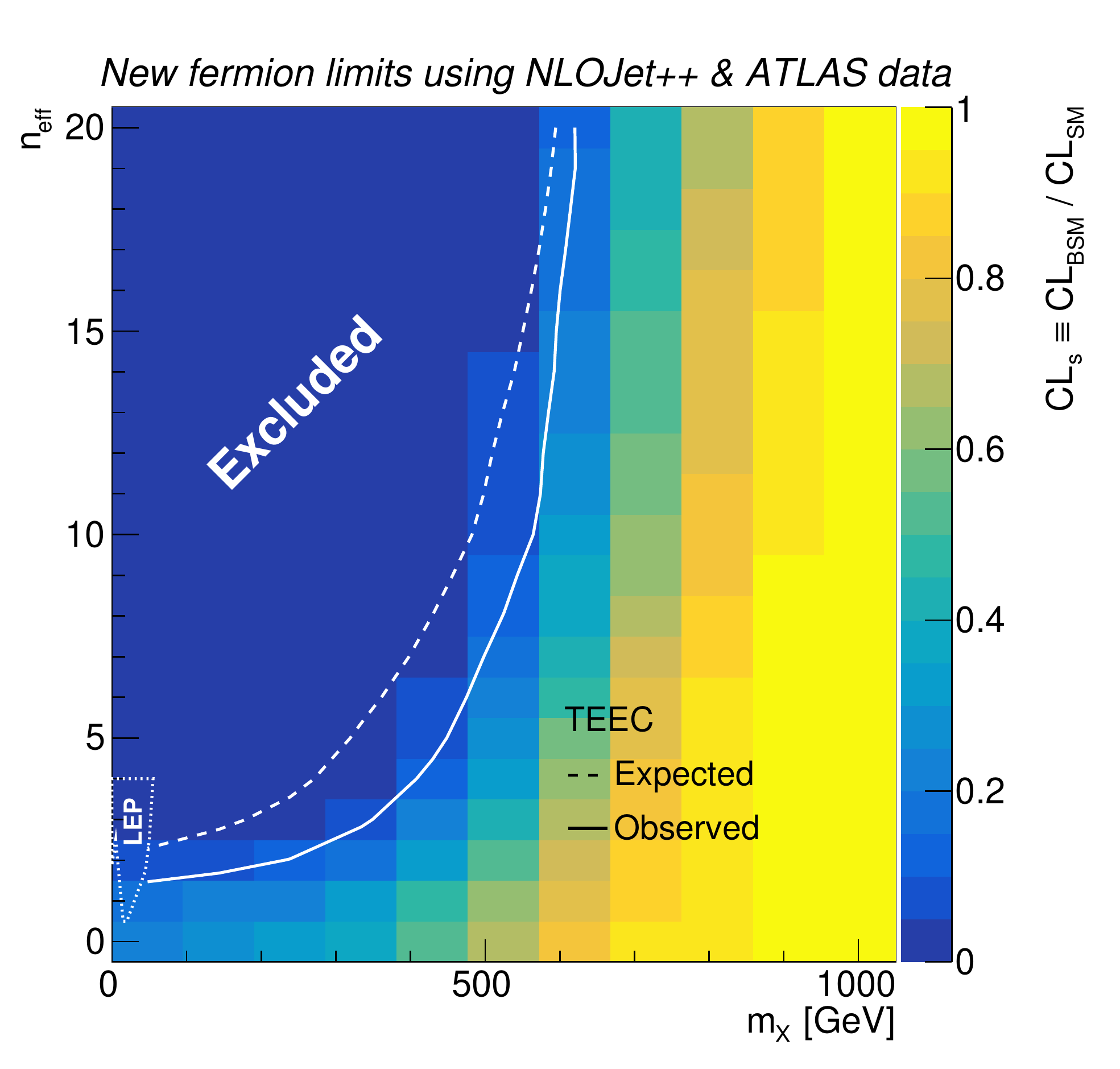}
\caption{The CL$_\mathrm{s}$ value calculated over the entire $n_\mathrm{eff}$, $m_X$ plane using the TEEC distributions.  A contour indicates where the CL$_\mathrm{s}$ surface crosses the 0.05 line for both the data (observed) and from the SM prediction (expected).  As is convention for direct searches, the surface is interpolated in the significance ($\sqrt{2}\times\mathrm{erf}^{-1}(1-2\mathrm{CL}_\mathrm{s})$) instead of the CL$_\mathrm{s}$ directly.  For reference, the LEP limits from Ref.~\cite{kaplan} are shown in the lower left corner.}
\label{fig:grid}
\end{figure}

\begin{figure}[H]
\centering
\includegraphics[width=0.8\textwidth]{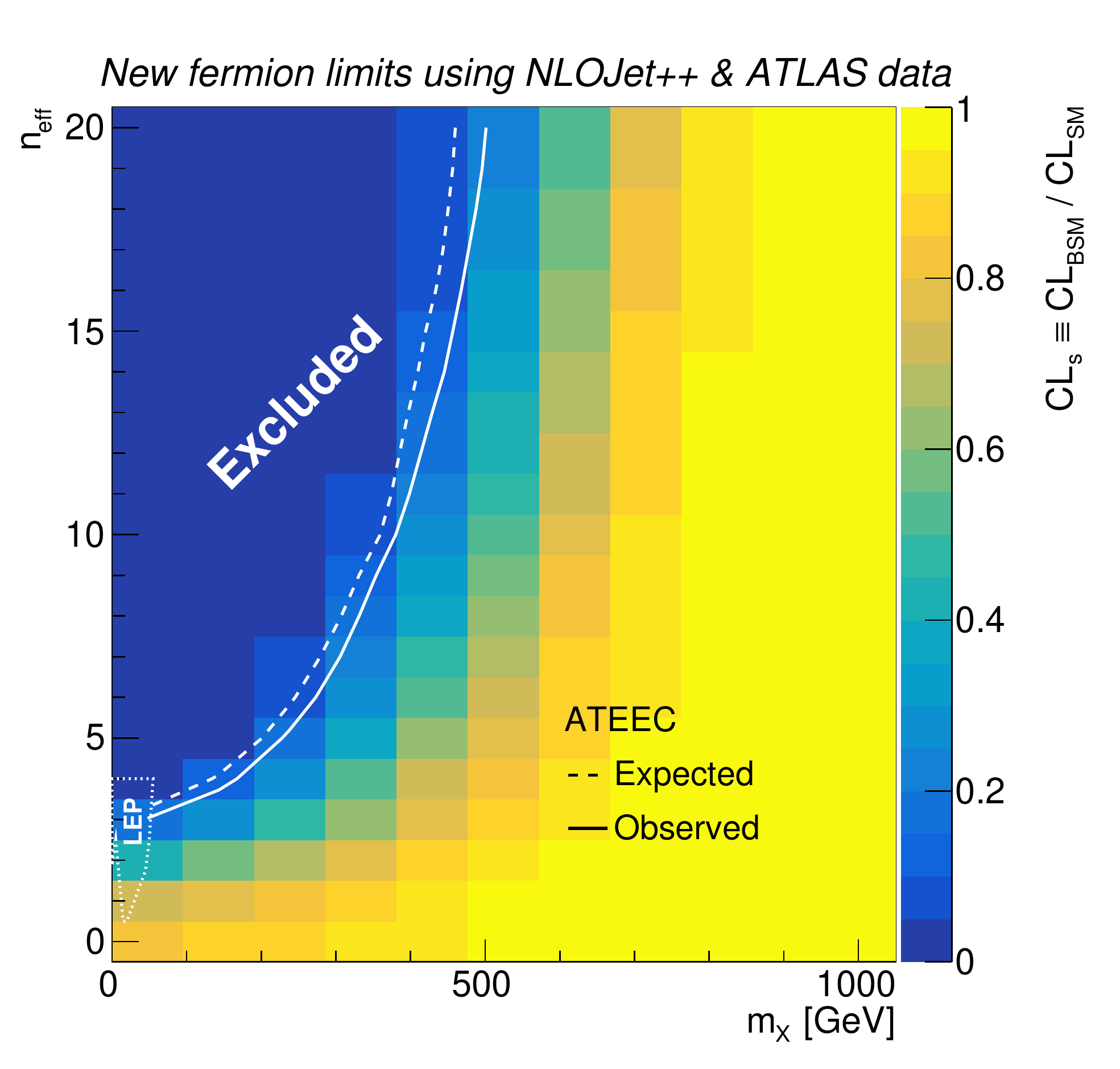}
\caption{The CL$_\mathrm{s}$ value calculated over the entire $n_\mathrm{eff}$, $m_X$ plane using the ATEEC distributions.  A contour indicates where the CL$_\mathrm{s}$ surface crosses the 0.05 line for both the data (observed) and from the SM prediction (expected).  As is convention for direct searches, the surface is interpolated in the significance ($\sqrt{2}\times\mathrm{erf}^{-1}(1-2\mathrm{CL}_\mathrm{s})$) instead of the CL$_\mathrm{s}$ directly.  For reference, the LEP limits from Ref.~\cite{kaplan} are shown in the lower left corner.}
\label{fig:gridA}
\end{figure}

Assuming the nominal factorization and renormalization scales, Figs.~\ref{fig:grid} and \ref{fig:gridA} summarize the exclusion limit over the entire $n_\mathrm{eff}$, $m_X$ plane, when using both the TEEC and ATEEC distributions. It is important to note that the ATEEC distributions lead to less powerful exclusion limits. This is understood to be due to the larger statistical uncertainties and the smaller number of degrees of freedom for the ATEEC with respect to the TEEC. Using the TEEC distributions, we are able to exclude masses up to 500 GeV for $n_\mathrm{eff}=20$ and for a nearly massless new fermion with $n_\mathrm{eff}=2$.

\clearpage

\section{Conclusions}
\label{sec:conclusions}

We have presented an interpretation of the recent ATLAS TEEC measurement in terms of constraints on new colored particles through their impact on the running of $\alpha_s$.  The MSSM with colored sparticle masses at 300 GeV is excluded using the nominal factorization and renormalization scale.  Including theoretical uncertainties can significantly vary the $p$-values, though the actual exclusion limits (in terms of $CL_\text{s}$) are not as affected.  Therefore, the future success of indirect searches for new colored particles will benefit from higher order calculations or new observables that are less sensitive to scale uncertainties.  Additionally, extending the measurement to higher $Q$ values will allow for higher new particle masses to be probed.  Indirect searches like the one we have presented are an important complement to direct searches because they are largely agnostic to the decay pattern of the BSM particles.  We may even find new particles hidden in the datasets already collected, waiting for increased precision in the future.

\section{Acknowledgements}

We would like to thank Ian Moult for helpful discussions during the early phase of this work and Matthias Schott and Bogdan Malaescu for their encouragement and Bogdan in particular for his feedback on the analysis. Communications with Fernando Barreiro are also thankfully acknowledged. This work was supported by the U.S.~Department of Energy, Office of Science under contract DE-AC02-05CH11231.

\clearpage

\bibliography{paper}

\end{document}